\documentclass[preprint,5p,times,twocolumn]{elsarticle}

\usepackage{lineno,hyperref}
\usepackage{multirow}
\usepackage{booktabs}
\usepackage{color}
\usepackage{amsmath}

\biboptions{sort&compress}

\makeatletter
\def\ps@pprintTitle{%
 \let\@oddhead\@empty
 \let\@evenhead\@empty
 \def\@oddfoot{}%
 \let\@evenfoot\@oddfoot}
\makeatother

\begin{document}

\begin{frontmatter}

\title{Compact LWFA-Based Extreme Ultraviolet Free Electron Laser: design constraints}



\author{Alexander~Molodozhentsev\corref{mycorrespondingauthor}}
\cortext[mycorrespondingauthor]{Corresponding author}
\ead{alexander.molodozhentsev@eli-beams.eu}

\author{Konstantin~O.~Kruchinin}
\address{Institute of Physics ASCR, v.v.i. (FZU), ELI-Beamlines, Za Radnici 835 Dolni Brezany 25241, Czech Republic}


\begin{abstract}
Combination of advanced high power laser technology, new acceleration methods and achievements in undulator development opens a way to build compact, high brilliance Free Electron Laser (FEL) driven by a laser wakefield accelerator (LWFA). Here we present a study outlining main requirements on the LWFA based Extreme Ultra Violet (EUV) FEL setup with the aim to reach saturation of photon pulse energy in a single unit commercially available undulator with the deflection parameter $K_0$ in a range of (1$\div$1.5). A dedicated electron beam transport which allows to control the electron beam slice parameters, including collective effects, required by the self-amplified spontaneous emission (SASE) FEL regime is proposed. Finally, a set of coherent photon radiation parameters achievable in the undulator section utilizing best experimentally demonstrated electron beam parameters are analyzed. As a result we demonstrate that the ultra-short (few fs level) pulse of the photon radiation with the wavelength in the EUV range can be obtained with the peak brilliance of $\sim$2$\times$10$^{30}$~photons/s/mm$^2$/mrad$^2$/0.1\%bw if the driver laser operates at the repetition rate of 25~Hz.
\end{abstract}


\end{frontmatter}

\section{Introduction}

In recent years linac-based FELs as a deliverer of coherent X-ray pulses changed the science landscape. Such facilities became an invaluable tool for the research in chemistry, biology, material science, medicine and physics~\cite{Pellegrini_Nature_2_2020}. Existing facilities (for example~\cite{XFEL-TDR, Swiss-FEL, Huang_NatPhot_6_2012}) broaden access to the technology, allowing more researchers to take advantage of the XFEL unique capabilities. Due to the fact that linac based XFELs utilize two well-established technologies: conventional or superconducting radio frequency (RF) accelerating structures and permanent magnet based undulators, such facilities have the total footprint from a
few hundred meters (Soft X-ray FELs) up to a few kilometers (Hard X-ray FELs) and, as a result, are extremely costly. Unique research importance and high cost leads to the situation when only a few such facilities operate around the world with a substantial lack of the user oriented beamtime.

The scientific success and tremendous demand from the photon beam user community stimulate intensive research to find competitive approaches which would lead to a significant size and cost reduction of the instruments. Constantly developing LWFA techniques make it very attractive candidate for the novel, compact FEL driver~\cite{Gruner-2007-TableFEL} especially in the light of recent progress in the laser technology~\cite{Mourou_OptPhotonNews_7_47_2011}, acceleration gain~\cite{Gonsalves_PhysRevLett_122_2019} and breakthroughs in electron beam quality improvement~\cite{Jalas-2021}. 

In this report we analyze combination of a LWFA electron beam parameters, demonstrated experimentally, with well established undulator technology in order to develop a beam transport design for a compact laser based EUV-FEL. We demonstrate that proposed setup is capable to generate high brightness coherent photon radiation reaching energy saturation only in a single unit planar undulator with the output photon brilliance comparable to the existing linac-based EUV-FEL facilities like FLASH (Germany)~\cite{Faatz-2017} and FERMI (Italy)~\cite{FERMI-2007}.

\section{Main constrains for a compact FEL}

Aiming to reach saturation of photon energy in one unit, first the undulator parameters must be fixed. For the purpose of this study we opted for a very well developed, commercially available undulator based on hybrid permanent magnet technology with the deflection parameter $K_0\sim$1 designed at ``Swiss-FEL''~\cite{Schmidt_FEL2012_THPD64_2012}. With this choice the undulator section length can be defined as no more than 4~m. Main undulator parameters are presented in Table~\ref{tab:und_params}.

From the SASE-FEL theory~\cite{Xie_NIMA_445_2000, Huang_PhysRevSTAB_10_2007} it is well known that the saturation length of the photon radiation energy is defined by the gain length of the FEL fundamental Gaussian mode and can be expressed as $L_{SAT}\approx20\times\left(1+\Delta\right) L_{1D}$. Here $L_{1D}$ is the gain length determined by one-dimensional FEL theory as a function of basic electron beam and undulator parameters. The $\Delta$-value represents a degradation of the SASE-FEL parameters caused by a finite energy spread, electron beam transverse size and diffraction effect. Taking into account realistic electron beam parameters, reachable in an electron beamline, one can estimate the $\Delta$ value to be around 1. Using this constrain one can conclude that in order to reach the photon energy saturation in a single-unit undulator with the length less than 4~m, the basic parameters of the electron beam have to be optimized to provide the 1D gain length no more than 0.1~m.

\begin{table*}[h!]
\caption{Summary of the undulator, electron beam and FEL parameters.}
\label{tab:und_params}
\centering
\begin{tabular}{lcclr}
\toprule
\textbf{Parameter}   & \textbf{Symbol}  &\textbf{Units}                &\multicolumn{2}{c}{\textbf{Value}} \\
\midrule
\\
\multicolumn{5}{c}{\textit{Undulator parameters}} \\
\\
Period               & $\lambda_u$      & mm &  15 & 15 \\
Gap                  & $g_u$              & mm & 4                       & 7           \\
Peak magnetic field  & $B_0$            & T  & 1                       & 0.54        \\
Undulator parameter & $K_0$            & -- & 1.4                     & 0.75        \\
\\
\multicolumn{5}{c}{\textit{LWFA-based electron beam parameters}} \\
\\
Energy & $W_k$            & MeV & 350 &350 \\
RMS emittance (slice) & $\varepsilon_n$ & $\pi$~mm~mrad & $<\varepsilon_{coh, n}$ & $<\varepsilon_{coh, n}$ \\
RMS beam size in undulator & $<\sigma_{x,y}>$ & $\mu$m & $\sim$ 25 & $\sim$ 20 \\
Energy spread (slice) & $\sigma_{\Delta\gamma/\gamma}$ & \% & 0.25 & 0.25 \\
RMS bunch length & $\sigma_z$ & $\mu$m & 1 &1 \\ 
Total bunch charge & $Q$ & pC & 25 & 35 \\
Peak current & $I_p$ & kA & 3 & 4.2 \\
\\
\multicolumn{5}{c}{\textit{LWFA-based FEL parameters}} \\
\\
Photon radiation wavelength & $\lambda_{ph, 1}$ & nm & 31.6 & 20.4 \\
Photon radiation energy & $E_{ph, 1}$ & eV & 39 & 60.5 \\
Coherent RMS emittance & $\varepsilon_{coh, n}$ & $\pi$~mm~mrad & 1.7 & 1.2 \\
1D Pierce parameter & $\rho_{1D}$ & -- & 0.0058 & 0.0065 \\
1D gain length & $L_{g, 1D}$ & m  & 0.12 & 0.1 \\
1D coherence length & $L_{coh, 1D}$ & $\mu$m & 0.4 & 0.25 \\
Total number of photons at saturation & $N_{photons}$ & -- & 6.2$\times$10$^{12}$ & 3.3$\times$10$^{12}$ \\
Relative FWHM frequency bandwidth & $\delta\lambda_{ph, 1}/\lambda_{ph, 1}$ & \% & 1.2 & 1.3 \\
Photon brilliance (1~Hz) & $B_{ph}$ & ph/s/mm$^2$/mrad$^2$/0.1\%bw & 2.6$\times$10$^{29}$ & 3.4$\times$10$^{29}$ \\
1D Peak power at saturation & $P_{ph, 1D}$ & GW & 5.4 & 5.2 \\
3D gain length & $L_{g, 3D}$ & m & 0.18 & 0.18 \\
3D total saturation length & $L_{sat, 3D}$ & m & $\sim$ 3.5 & $\sim$ 3.5 \\
\bottomrule
\end{tabular}
\end{table*}

With the above considerations in mind an analytic analysis using 1D and 3D SASE-FEL models has been performed assuming Gaussian distribution of electrons in the bunch. Main constraints on LWFA electron beam and corresponding coherent photon radiation parameters have been derived and summarized in Table~\ref{tab:und_params}.
The results indicate that in order to keep the saturation length of around 3.5~m the electron beam energy have to be around 350~MeV. Such energy with stable and reproducible electron beam parameters has been demonstrated experimentally in a compact laser-plasma interaction region~\cite{Maier_PhysRevX_10_2020}, utilizing a moderate power of the compressed laser pulse.

Experimental results indicate that the electron beam with sufficient bunch charge and the energy required to operate the FEL in the EUV range 
can be obtained using a ``Joule-class'' laser with a pulse duration of $\sim$30~fs. Such lasers are capable to run with a repetition rate of up to 50 Hz~\cite{Tyler_DUHA} which can boost the peak brilliance of the photon radiation to values comparable with the existing RF based soft X-ray FELs. 

\section{Laser-plasma acceleration for EUV-FEL}

\begin{figure*}[t!]
\includegraphics[width=\textwidth]{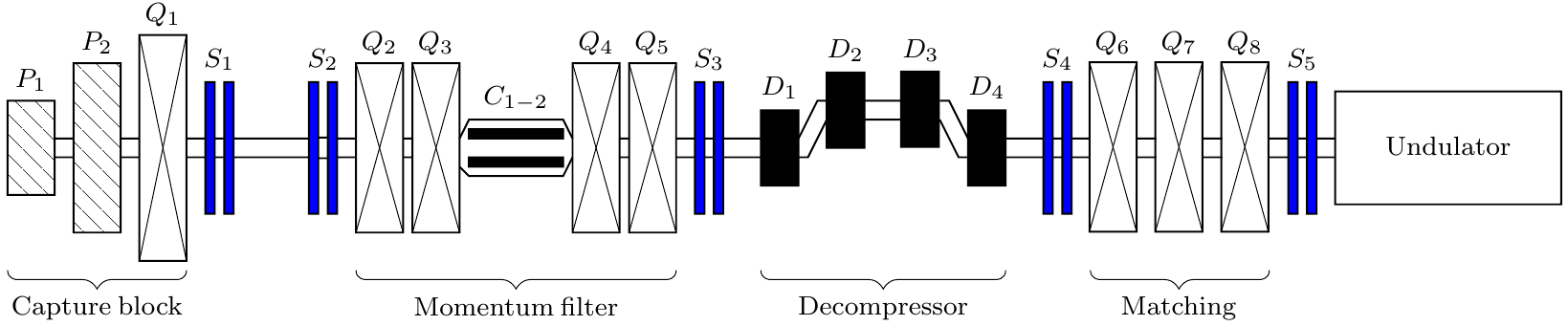}
\caption{\label{fig:beamline_sch}Schematic diagram representing proposed beamline. Here $P_1$ and $P_2$ are permanent quadrupole magnets, $Q_1\div Q_{8}$ are electromagnetic quadrupoles, $C_{1-2}$ is a pair of horizontal and vertical collimators, $S_1\div S_5$ are pairs of steering magnets (vertical and horizontal) and $D_1\div D_4$ are dipole magnets of the decompressor chicane.}
\end{figure*}

Plasma-based accelerators are of great interest because of their ability to sustain extremely large acceleration gradients. The electric field in plasma of the order of 100~GV/m has been demonstrated experimentally~\cite{Malka-2002, Hafz-2008}. Consequently, the accelerating structure required to produce LWFA electron beam with the energy around (300$\div$400)~MeV will have the length of a few centimeters. The laser-plasma acceleration technology has made a great progress recently to provide a stable operation and produce the electron beam with peak current of a few kA and energy in a GeV range.

Analyzing published experimental results, obtained by different groups~\cite{Assmann_Eur.Phys.Journ.Spec_24_2020}, one can identify typical parameters of the electron bunch expected from the LWFA as following: (1) the normalized transverse RMS emittance of the electron beam is in the range of (0.2$\div$0.5)~$\pi$~$\mu$m~rad for the electron beam with the peak energy of up to 1~GeV; (2) the RMS transverse divergence is in range of (0.5$\div$ 1)~mrad; (3) the RMS energy spread with the FWHM bunch length of a few
$\mu$m can be reduced down to 1~\% keeping the bunch charge of a few tens of pC. Experimentally demonstrated set of the LWFA electron bunch initial parameters fits well the SASE-FEL requirements assuming that one can preserve the electron beam quality while transporting electrons from the LWFA source to the undulator.

\section{Electron beam transport for EUV-FEL}

In order to deliver the LWFA electron beam to an FEL undulator a dedicated beam transport has to be designed to provide: capture of electrons from the LWFA source~\cite{Hofmann_PhysRevSTAB_16_2013}; effective transport of the electron beam with preservation of its quality; possible manipulation with the electron bunch in the longitudinal phase-space~\cite{Maier_PhysRevX_2_2012} and, finally, matching of the electron beam to the undulator~\cite{Loulergue_N.Journ.Physics_17_2015}. Moreover,  a combination of the significant initial transverse divergence  and the large energy spread of LWFA electrons leads to an intrinsic growth of the normalized transverse beam emittance in a drift space after the laser plasma interaction area~\cite{Migliorati_PhysRevSTAB_16_2013}. Therefore, the beam transport has to be designed in such a way to mitigate the emittance growth due to the chromatic effects. Apart from chromatic effects the laser driven electron beam,  having the bunch duration of a few fs and the bunch charge of a few tens of pC, will suffer from collective effects which have to be analysed and taken into account in order to prevent the beam emittance degradation.

The conceptual solution for such beam transport which allows one to reach the SASE-EUV-FEL regime in a single unit undulator is presented in Fig.~\ref{fig:beamline_sch}. The proposed beamline consists of the following elements: a capture block, a momentum filter, a decompressor chicane and a matching block.

The capture block consists of a pair Halbach-type permanent quadrupole magnets (PQM)~\cite{Halbach_NIM_169_1980} and one conventional electromagnetic quadrupole (EMQ). In spite of lack in tuning flexibility PQMs provide  required magnetic field gradient for capturing highly divergent electron beam which is unreachable by conventional EMQs. One possible solution to improve the flexibility would be the use of active plasma lens~\cite{Panofsky_1950} instead of the PQM pair, however this option needs to be further investigated. As a unit, the capture block has been designed to prevent growth of the normalized RMS transverse emittance and, at the same time, to focus the electron beam in both planes to propagate through a significant drift space required to separate the driver laser and electron beam after the interaction point.

The momentum filter, consisting of four EMQs with integrated pair of vertical and horizontal collimators, is placed after the capture block. The main purpose of this device is to allow propagation further down the beamline for those electrons with the desired reference energy only~\cite{Molodozhentsev_FLS2018_TUA2WC02_2018}. After passing the first triplet of quadrupoles in the capture block, due to the chromatic aberrations and significant initial energy spread, the electron beam will form a halo consisting of particles with energies different from the reference. Inside the momentum filter the halo particles are effectively filtered out by collimators. As a result, the projected transverse RMS normalized emittance can be controlled in addition to a decompressor chicane, which is the next block of the proposed beamline.

\begin{figure}[h!]
\includegraphics[width=\columnwidth]{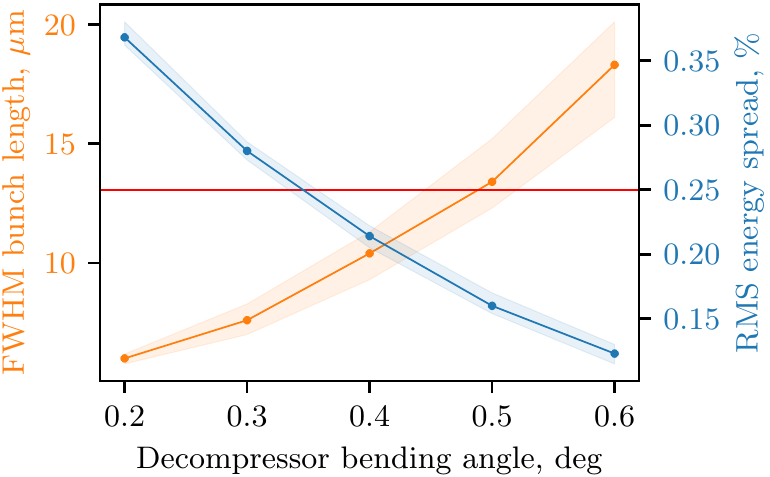}
\caption{\label{fig:decompressor_optimization} FWHM bunch length (orange) and slice RMS energy spread (blue) as a function of the decompressor bending angle. The solid lines represent the case for the bunch charge of 35~pC. Shaded areas represent the bunch charge variation in the range of (25$\div$45)~pC.}
\end{figure}

As it was mentioned before the LWFA electron beam has a relatively large (compared to the conventional RF accelerators) initial energy spread which makes it impossible to reach FEL performance. As proposed in~\cite{Maier_PhysRevX_2_2012} this problem can be solved by introducing a dispersive section (a decompressor chicane) into the beamline where the electron bunch is stretched longitudinally and effectively sorted by energy resulting in a reduction of the local (slice) energy spread at the cost of a reduced peak current and an energy chirp. The slice energy spread in this case can be controlled by adjusting the bending angle of the chicane dipole magnets. 
For a proper optimization of the magnetic chicane performance the space charge effect at low electron energy has to be taken into account. The space charge force will change the transverse and longitudinal particle distribution in the bunch, leading to almost uniform longitudinal beam profile. 

\begin{figure}[h!]
\includegraphics[width=\columnwidth]{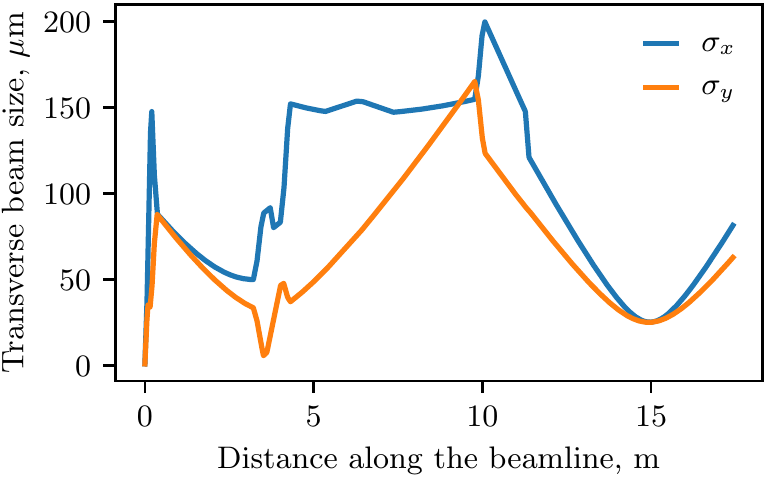}
\caption{\label{fig:sigma_beamline} Electron beam envelope (horizontal and vertical) along the beam transport. Parameters: electron energy is 350~MeV, decompressor bending angle is 0.35~deg, initial bunch charge is 45~pC. The center of the single unit undulator is located at 15~m from the source.}
\end{figure}

The result of such optimization is presented in Fig.~\ref{fig:decompressor_optimization} showing variation of the FWHM bunch length and the average slice RMS energy spread as a function of the decompressor bending angle for various bunch charges. The slice length is determined by the coherence length for the SASE-FEL regime. The required slice energy spread is defined by the SASE-FEL constraints discussed above. In order to keep the slice energy spread below 0.25~\% dictated by the FEL study performed in the previous chapter the bending angle of the decompressor have to be no less than 0.35~degree and the initial projected RMS energy spread of the electron bunch has to be 0.5~\%. Further increase of the angle will lead to the peak current reduction. To keep the current at the same level more charge is required. Increasing the charge will intensify collective effects which, at some point, will lead to the degradation of the electron beam quality~\cite{Molodozhentsev_IPAC2018_THPAF033_2018}.

Finally, the matching block consisting of three EMQs is placed before the undulator. This block allows to focus the electron beam to the required transverse dimension, both vertically and horizontally, in the middle of the undulator. 

In addition to the described beam transport sections enough drift space have to be reserved between blocks in order to put steering magnets for the orbit correction and various beam diagnostic devices~\cite{Kruchinin_IBIC2019_MOPP035_2019}. Additional collimator can be placed in front of the undulator entrance to clean the beam halo and control the slice normalized transverse emittance, required by the SASE-FEL regime.

The electron beam dynamics for the proposed beamline has been studied using TraceWin code~\cite{Uriot_IPAC2015_MOPWA008_2015} taken into account the space charge effect in combination with the collimation of the beam. Initial electron beam parameters were optimized to provide the required slice electron beam parameters presented in Table~1. An example of the transverse beam size variation along the beamline is presented in Fig.~\ref{fig:sigma_beamline}. In this case the initial projected RMS energy spread is equal to 0.5~\%, the RMS transverse divergence of the electron beam is 0.5~mrad, the RMS bunch length is 1~$\mu m$ and the total bunch charge is 40~pC. As one can see the proposed beam transport design allows to focus the beam down to $\sim$25~$\mu m$ in both planes with propagation efficiency of about 80~\% keeping the normalized RMS projected emittance at the level of 0.6~$\pi$~mm~mrad at the entrance of the undulator. Proper optimization of the decompressor chicane allows to keep the slice energy spread below 0.25~\%  and the slice RMS normalized transverse emittance no more than 0.4~$\pi$~mm~mrad, as required for reaching the FEL performance.

\section{EUV-FEL regime}

The start-to-end simulations of the electron beam dynamics with collective effects and the generation of the coherent photon radiation in the chosen undulator have been performed by using the optimized set of the decompressor chicane. The SASE-FEL regime in the case of a single-unit undulator with the gap size of 4~mm has been studied to confirm the saturation of the radiation power at the exit of the undulator including an external seeding to improve the longitudinal coherence of the generated photon pulse. 

\begin{figure}[h!]
\includegraphics[width=\columnwidth]{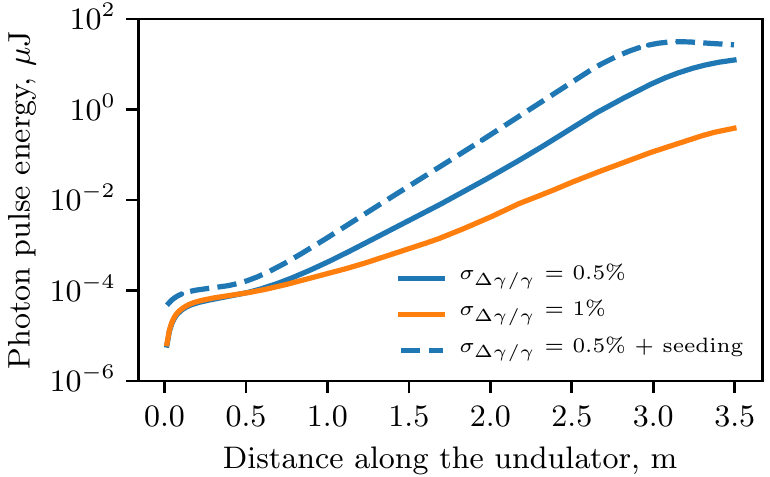}
\caption{\label{fig:und_photon_rad} Photon pulse energy amplification for different projected energy spread in the case of the optimized decompressor angle.}
\end{figure}

\begin{figure}[htb!]
\includegraphics[width=\columnwidth]{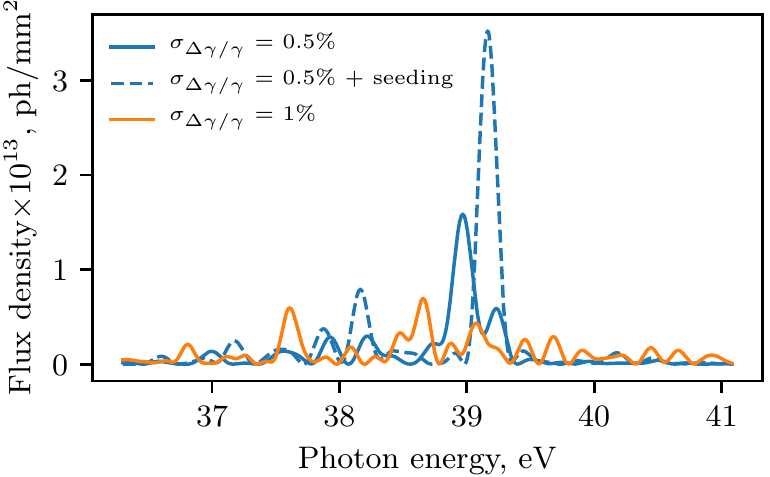}
\caption{\label{fig:photon_rad_spectrum} Flux density in units of number of photons/pulse/mm$^2$ per the energy bandwidth of 0.1\% without (solid line) and with an external seeding (dashed line). }
\end{figure}

The results of the study are presented in Fig.~\ref{fig:und_photon_rad} and Fig.~\ref{fig:photon_rad_spectrum} showing photon energy amplification inside the undulator and corresponding energy spectrum respectively. As one can notice the saturation of the photon pulse energy can be reached after approximately 3.5~m if the initial projected RMS energy spread of the laser-driven electron beam is 0.5~\%. The corresponding value of the radiation peak brilliance is 7.5$\times$10$^{28}$~photons/pulse/mrad$^2$/mm$^2$/0.1\%bw. If the initial projected RMS energy spread is increased to 1~\% the saturation of the photon pulse energy along the same undulator can not be obtained.

An external seeding with proper parameters allows to reduce the saturation length and increase the photon pulse energy. To demonstrate this an external signal with the wavelength of 31.6~nm  and the power of 1.2~kW was applied to the case of 0.5~\% initial energy spread. The photon radiation energy saturation in this case can be reached after around 3~meters producing a single spike energy spectrum.

The photon radiation wavelength can be tuned by varying the undulator gap. An example output parameters for the gap size of 7~mm are presented in Table~\ref{tab:und_params}. It must be taken into account that for each gap position the electron peak current and consequently the beamline parameters have to be re-optimized in order to reach saturation of photon energy at the fixed undulator length. 

\section{Conclusion}

A comprehensive analysis of the LWFA electron beam main parameters has been performed in order to provide the saturation of the photon energy in the single-unit ``Swiss-FEL'' type undulator. It was shown that the LWFA electron beam with the energy of 350~MeV is capable of generating a fundamental harmonic photon radiation with the wavelength in the range of (20.4$\div$31.6)~nm depending on the undulator gap size. A dedicated electron beam transport has been presented, allowing to deliver the electron beam from the LWFA source to the FEL undulator with the parameters needed to reach saturation of the photon radiation power in the EUV  wavelength range in the single undulator unit with the length of less than 4~m.

Is was demonstrated that the saturation of the photon radiation power can be reached at the end of the single unit ``Swiss-FEL'' type undulator even without an additional external seeding after optimization of the electron beam slice parameters. The proper seeding signal allows to reduce the saturation length and generate the photon energy spike with the improved longitudinal coherence. The peak brilliance of the single-spike photon radiation with the wavelength of $\sim$31.6~nm was found to be $\sim$2$\times$10$^{30}$~photon/s/mrad$^2$/mm$^2$/0.1\%bw in the case of 25~Hz repetition rate of the driver laser, which makes it comparable with the linac-based shoft X-ray FELs.

\section{Acknowledgement}

This work has been supported by the project ``Advanced Research using High Intensity Laser produced Photons and Particles'' (ADONIS) (CZ.02.1.01/0.0/0.0/16019/0000789) from European Regional Development Fund (ERDF)
and by ``High Field Initiative'' (HiFI) project (CZ.02.1.01/0.0/0.0/15-003/0000449).

Authors thank to Prof.~S.V.~Bulanov and Dr.~A.R.~Maier,  for many useful discussions and to Dr.~G.~Korn for his support of this activity at ELI-Beamlines.

\bibliography{LUIS_EUV_FEL}

\begin{thebibliography}{10}
\expandafter\ifx\csname url\endcsname\relax
  \def\url#1{\texttt{#1}}\fi
\expandafter\ifx\csname urlprefix\endcsname\relax\def\urlprefix{URL }\fi
\expandafter\ifx\csname href\endcsname\relax
  \def\href#1#2{#2} \def\path#1{#1}\fi

\bibitem{Pellegrini_Nature_2_2020}
C.~Pellegrini, \href{https://doi.org/10.1038/s42254-020-0197-1}{{The
  development of XFELs}}, Nature Reviews Physics 2~(7) (2020) 330--331 (2020).
\newblock \href {https://doi.org/10.1038/s42254-020-0197-1}
  {\path{doi:10.1038/s42254-020-0197-1}}.
\newline\urlprefix\url{https://doi.org/10.1038/s42254-020-0197-1}

\bibitem{XFEL-TDR}
{XFEL: The European X-Ray Free-Electron Laser. Technical design report} (7
  2006).
\newblock \href {https://doi.org/10.3204/DESY_06-097}
  {\path{doi:10.3204/DESY_06-097}}.

\bibitem{Swiss-FEL}
C.~J. Milne, T.~Schietinger, M.~Aiba, et. al.,
  \href{https://www.mdpi.com/2076-3417/7/7/720}{{SwissFEL: The Swiss X-ray Free
  Electron Laser}}, Applied Sciences 7~(7) (2017).
\newblock \href {https://doi.org/10.3390/app7070720}
  {\path{doi:10.3390/app7070720}}.
\newline\urlprefix\url{https://www.mdpi.com/2076-3417/7/7/720}

\bibitem{Huang_NatPhot_6_2012}
Z.~Huang, I.~Lindau, \href{https://doi.org/10.1038/nphoton.2012.184}{{SACLA}
  hard-{X}-ray compact {FEL}}, Nature Photonics 6~(8) (2012) 505--506 (Aug
  2012).
\newblock \href {https://doi.org/10.1038/nphoton.2012.184}
  {\path{doi:10.1038/nphoton.2012.184}}.
\newline\urlprefix\url{https://doi.org/10.1038/nphoton.2012.184}

\bibitem{Gruner-2007-TableFEL}
F.~Gruner, S.~Becker, U.~Schramm, et. al, Design considerations for table-top,
  laser-based {VUV} and {X}-ray free electron lasers, Appl. Phys. B 86 (2007)
  431--435 (2007).
\newblock \href {https://doi.org/https://doi.org/10.1007/s00340-006-2565-7}
  {\path{doi:https://doi.org/10.1007/s00340-006-2565-7}}.

\bibitem{Mourou_OptPhotonNews_7_47_2011}
G.~Mourou, T.~Tajima,
  \href{http://www.osa-opn.org/abstract.cfm?URI=opn-22-7-47}{The extreme light
  infrastructure: Optics' next horizon}, Opt. Photon. News 22~(7) (2011) 47--51
  (Jul 2011).
\newblock \href {https://doi.org/10.1364/OPN.22.7.000047}
  {\path{doi:10.1364/OPN.22.7.000047}}.
\newline\urlprefix\url{http://www.osa-opn.org/abstract.cfm?URI=opn-22-7-47}

\bibitem{Gonsalves_PhysRevLett_122_2019}
A.~J. Gonsalves, K.~Nakamura, J.~Daniels, {et.~al.},
  \href{https://link.aps.org/doi/10.1103/PhysRevLett.122.084801}{Petawatt laser
  guiding and electron beam acceleration to 8 gev in a laser-heated capillary
  discharge waveguide}, Phys. Rev. Lett. 122 (2019) 084801 (Feb 2019).
\newblock \href {https://doi.org/10.1103/PhysRevLett.122.084801}
  {\path{doi:10.1103/PhysRevLett.122.084801}}.
\newline\urlprefix\url{https://link.aps.org/doi/10.1103/PhysRevLett.122.084801}

\bibitem{Jalas-2021}
S.~Jalas, M.~Kirchen, P.~Messner, et. al, {B}ayesian {O}ptimization of a
  {L}aser-{P}lasma {A}ccelerator, Phy. Rev. Lett. 126~(104801) (2021) 505--506
  (2021).
\newblock \href
  {https://doi.org/https://doi.org/10.1103/PhysRevLett.126.104801}
  {\path{doi:https://doi.org/10.1103/PhysRevLett.126.104801}}.

\bibitem{Faatz-2017}
B.~Faatz, M.~Braune, O.~Hensler, et. al, The {FLASH} {F}acility: Advanced
  {O}ptions for {FLASH2} and {F}uture {P}erspectives, Appl. Sci. 7~(1114)
  (2017).
\newblock \href {https://doi.org/https://doi.org/10.3390/app7111114}
  {\path{doi:https://doi.org/10.3390/app7111114}}.

\bibitem{FERMI-2007}
C.~Bocchetta, et. al, Conceptual {D}esign {R}eport for the {FERMI@E}lettra
  project, ST/F-TN-07/12 (2007).

\bibitem{Schmidt_FEL2012_THPD64_2012}
T.~Schmidt, P.~B\"ohler, M.~Br\"ugger, {et.~al.}, {SwissFEL} {U15} prototype
  design and first results, in: Proc. IBIC'12, JACoW Publishing, Geneva,
  Switzerland, 2012, pp. 666 -- 669 (2012).

\bibitem{Xie_NIMA_445_2000}
M.~Xie,
  \href{http://www.sciencedirect.com/science/article/pii/S0168900200001145}{Exact
  and variational solutions of 3d eigenmodes in high gain fels}, Nucl. Instrum.
  Methods Phys. Res., A 445~(1) (2000) 59 -- 66 (2000).
\newblock \href {https://doi.org/https://doi.org/10.1016/S0168-9002(00)00114-5}
  {\path{doi:https://doi.org/10.1016/S0168-9002(00)00114-5}}.
\newline\urlprefix\url{http://www.sciencedirect.com/science/article/pii/S0168900200001145}

\bibitem{Huang_PhysRevSTAB_10_2007}
Z.~Huang, K.-J. Kim,
  \href{https://link.aps.org/doi/10.1103/PhysRevSTAB.10.034801}{Review of x-ray
  free-electron laser theory}, Phys. Rev. ST Accel. Beams 10 (2007) 034801 (Mar
  2007).
\newblock \href {https://doi.org/10.1103/PhysRevSTAB.10.034801}
  {\path{doi:10.1103/PhysRevSTAB.10.034801}}.
\newline\urlprefix\url{https://link.aps.org/doi/10.1103/PhysRevSTAB.10.034801}

\bibitem{Maier_PhysRevX_10_2020}
A.~R. Maier, N.~M. Delbos, T.~Eichner, {et.~al.},
  \href{https://link.aps.org/doi/10.1103/PhysRevX.10.031039}{Decoding sources
  of energy variability in a laser-plasma accelerator}, Phys. Rev. X 10 (2020)
  031039 (Aug 2020).
\newblock \href {https://doi.org/10.1103/PhysRevX.10.031039}
  {\path{doi:10.1103/PhysRevX.10.031039}}.
\newline\urlprefix\url{https://link.aps.org/doi/10.1103/PhysRevX.10.031039}

\bibitem{Tyler_DUHA}
T.~Green, R.~Antipenkov, P.~Bakule, et. al, {L2-DUHA 100TW High Repetition Rate
  Laser System at ELI-Beamlines: Key Design Considerations}, The Review of
  Laser Engineering 49~(2) (2021).

\bibitem{Malka-2002}
V.~Malka, S.~Fritzler, E.~Lefebvre, et. al, Electron {A}cceleration by a {W}ake
  {F}ield {F}orced by an {I}ntense {U}ltrashort {L}aser {P}ulse, Science 298
  (2002) 1596--1600 (November 2002).

\bibitem{Hafz-2008}
N.~Hafz, T.~Jeong, W.~Il, et. al, Electron stable generation of {G}e{V}-class
  electron beams from self-guided laser-plasma channels, Nature (2008) 571--577
  (August 2008).

\bibitem{Assmann_Eur.Phys.Journ.Spec_24_2020}
R.~W. Assmann, M.~K. Weikum, T.~Akhter, {et.~al.},
  \href{https://doi.org/10.1140/epjst/e2020-000127-8}{{EuPRAXIA Conceptual
  Design Report}}, The European Physical Journal Special Topics 229~(24) (2020)
  3675--4284 (2020).
\newblock \href {https://doi.org/10.1140/epjst/e2020-000127-8}
  {\path{doi:10.1140/epjst/e2020-000127-8}}.
\newline\urlprefix\url{https://doi.org/10.1140/epjst/e2020-000127-8}

\bibitem{Hofmann_PhysRevSTAB_16_2013}
I.~Hofmann, \href{https://link.aps.org/doi/10.1103/PhysRevSTAB.16.084201}{Halo
  coupling and cleaning by a space charge resonance in high intensity beams},
  Phys. Rev. ST Accel. Beams 16 (2013) 084201 (Aug 2013).
\newblock \href {https://doi.org/10.1103/PhysRevSTAB.16.084201}
  {\path{doi:10.1103/PhysRevSTAB.16.084201}}.
\newline\urlprefix\url{https://link.aps.org/doi/10.1103/PhysRevSTAB.16.084201}

\bibitem{Maier_PhysRevX_2_2012}
A.~R. Maier, A.~Meseck, S.~Reiche, {et.~al.},
  \href{https://link.aps.org/doi/10.1103/PhysRevX.2.031019}{Demonstration
  scheme for a laser-plasma-driven free-electron laser}, Phys. Rev. X 2 (2012)
  031019 (Sep 2012).
\newblock \href {https://doi.org/10.1103/PhysRevX.2.031019}
  {\path{doi:10.1103/PhysRevX.2.031019}}.
\newline\urlprefix\url{https://link.aps.org/doi/10.1103/PhysRevX.2.031019}

\bibitem{Loulergue_N.Journ.Physics_17_2015}
A.~Loulergue, M.~Labat, C.~Evain, {et.~al.},
  \href{https://doi.org/10.1088/1367-2630/17/2/023028}{Beam manipulation for
  compact laser wakefield accelerator based free-electron lasers}, New Journal
  of Physics 17~(2) (2015) 023028 (feb 2015).
\newblock \href {https://doi.org/10.1088/1367-2630/17/2/023028}
  {\path{doi:10.1088/1367-2630/17/2/023028}}.
\newline\urlprefix\url{https://doi.org/10.1088/1367-2630/17/2/023028}

\bibitem{Migliorati_PhysRevSTAB_16_2013}
M.~Migliorati, A.~Bacci, C.~Benedetti, {et.~al.},
  \href{https://link.aps.org/doi/10.1103/PhysRevSTAB.16.011302}{Intrinsic
  normalized emittance growth in laser-driven electron accelerators}, Phys.
  Rev. ST Accel. Beams 16 (2013) 011302 (Jan 2013).
\newblock \href {https://doi.org/10.1103/PhysRevSTAB.16.011302}
  {\path{doi:10.1103/PhysRevSTAB.16.011302}}.
\newline\urlprefix\url{https://link.aps.org/doi/10.1103/PhysRevSTAB.16.011302}

\bibitem{Halbach_NIM_169_1980}
K.~Halbach,
  \href{http://www.sciencedirect.com/science/article/pii/0029554X80900944}{Design
  of permanent multipole magnets with oriented rare earth cobalt material},
  Nuclear Instruments and Methods 169~(1) (1980) 1 -- 10 (1980).
\newblock \href {https://doi.org/https://doi.org/10.1016/0029-554X(80)90094-4}
  {\path{doi:https://doi.org/10.1016/0029-554X(80)90094-4}}.
\newline\urlprefix\url{http://www.sciencedirect.com/science/article/pii/0029554X80900944}

\bibitem{Panofsky_1950}
W.~K.~H. Panofsky, W.~R. Baker, A focusing device for the external 350-{MeV}
  proton beam of the 184-inch cyclotron at berkeley, Rev. Sci. Instrum. 21
  (1950).
\newblock \href {https://doi.org/https://10.1063/1.1745611}
  {\path{doi:https://10.1063/1.1745611}}.

\bibitem{Molodozhentsev_FLS2018_TUA2WC02_2018}
A.~Molodozhentsev, G.~Korn, A.~Maier, {et.~al.},
  \href{http://jacow.org/fls2018/papers/tua2wc02.pdf}{"{LWFA-}driven"{} {F}ree
  {E}lectron {L}aser for {ELI-B}eamlines}, in: Proc. 60th ICFA Advanced Beam
  Dynamics Workshop (FLS'18), Shanghai, China, 5-9 March 2018, no.~60 in ICFA
  Advanced Beam Dynamics Workshop, JACoW Publishing, Geneva, Switzerland, 2018,
  pp. 62--67, https://doi.org/10.18429/JACoW-FLS2018-TUA2WC02 (June 2018).
\newblock \href {https://doi.org/doi:10.18429/JACoW-FLS2018-TUA2WC02}
  {\path{doi:doi:10.18429/JACoW-FLS2018-TUA2WC02}}.
\newline\urlprefix\url{http://jacow.org/fls2018/papers/tua2wc02.pdf}

\bibitem{Molodozhentsev_IPAC2018_THPAF033_2018}
A.~Molodozhentsev, K.~Kruchinin, L.~Pribyl,
  \href{http://jacow.org/ipac2018/papers/thpaf033.pdf}{{D}egradation of
  {E}lectron {B}eam {Q}uality for a {C}ompact {L}aser{-B}ased {FEL}}, in: Proc.
  9th International Particle Accelerator Conference (IPAC'18), Vancouver, BC,
  Canada, April 29-May 4, 2018, no.~9 in International Particle Accelerator
  Conference, JACoW Publishing, Geneva, Switzerland, 2018, pp. 3029--3031,
  https://doi.org/10.18429/JACoW-IPAC2018-THPAF033 (June 2018).
\newblock \href {https://doi.org/doi:10.18429/JACoW-IPAC2018-THPAF033}
  {\path{doi:doi:10.18429/JACoW-IPAC2018-THPAF033}}.
\newline\urlprefix\url{http://jacow.org/ipac2018/papers/thpaf033.pdf}

\bibitem{Kruchinin_IBIC2019_MOPP035_2019}
K.~Kruchinin, D.~Kocon, A.~Lyapin, {et.~al.},
  \href{http://jacow.org/ibic2019/papers/mopp035.pdf}{{Electron Beam
  Diagnostics Concept for the LWFA Driven FEL at ELI-Beamlines}}, in: Proc.
  IBIC'19, no.~8 in International Beam Instrumentation Conferenc, JACoW
  Publishing, Geneva, Switzerland, 2019, pp. 184--187,
  https://doi.org/10.18429/JACoW-IBIC2019-MOPP035 (nov 2019).
\newblock \href {https://doi.org/10.18429/JACoW-IBIC2019-MOPP035}
  {\path{doi:10.18429/JACoW-IBIC2019-MOPP035}}.
\newline\urlprefix\url{http://jacow.org/ibic2019/papers/mopp035.pdf}

\bibitem{Uriot_IPAC2015_MOPWA008_2015}
D.~Uriot, N.~Pichoff,
  \href{http://jacow.org/ipac2015/papers/mopwa008.pdf}{{S}tatus of {T}race{W}in
  {C}ode}, in: Proc. 6th International Particle Accelerator Conference
  (IPAC'15), Richmond, VA, USA, May 3-8, 2015, no.~6 in International Particle
  Accelerator Conference, JACoW, Geneva, Switzerland, 2015, pp. 92--94,
  https://doi.org/10.18429/JACoW-IPAC2015-MOPWA008 (June 2015).
\newblock \href
  {https://doi.org/https://doi.org/10.18429/JACoW-IPAC2015-MOPWA008}
  {\path{doi:https://doi.org/10.18429/JACoW-IPAC2015-MOPWA008}}.
\newline\urlprefix\url{http://jacow.org/ipac2015/papers/mopwa008.pdf}

\end{thebibliography}

\end{document}